\documentstyle[12pt,preprint,aps,prd]{revtex}

\tightenlines

\def\text{\mbox}
\def\ll{\left\langle}
\def\rr{\right\rangle}
\def\[{\begin{eqnarray}}
\def\]{\end{eqnarray}}

\begin{document}

\title {GAUGE--FIXING INDEPENDENCE OF TEST FIELDS IN YANG--MILLS THEORIES}
    
\author{Sergio M.\ Iguri\thanks{siguri@df.uba.ar} and Francisco D.
\ Mazzitelli\thanks{fmazzi@df.uba.ar}}

\address{Departamento de F\'{\i}sica {\it J.\ J.\ Giambiagi}, Facultad de 
Ciencias Exactas y Naturales \\ Universidad de Buenos Aires - Ciudad 
Universitaria, Pabell\'{o}n I, 1428 Buenos Aires, Argentina}

\maketitle

\begin{abstract}
We derive a generalized Nielsen identity for the
case of Yang-Mills theories that include some classical fields. We discuss 
under which circumstances the effective action of the classical fields
(i.e., after integration of quantum fields) becomes gauge--fixing
independent. We conclude that classical test fields provide a 
physical insight into the problem of the gauge--fixing dependence of
the quantum effective action. 
\end{abstract}
\bigskip


\section{INTRODUCTION}
\label{Introduction}
\bigskip

As it is well known, whatever the scheme employed to quantize a 
gauge theory 
is (i.e., Gupta-Bleuler, Fadeev--Popov, BRST, Batalin--Vilkovisky methods), it is 
necessary to fix a particular gauge in order to carry on the quantization 
program. Gauge--fixing is implemented introducing into the classical 
action an additional non--invariant term, the so called gauge--breaking 
term. The resulting classical effective action is a gauge--parameter 
dependent functional on the field--configuration manifold. Required gauge 
invariance of the quantum theory follows, in turn, from the fact that 
expectation values of physical magnitudes become independent of the choice of 
this particular term.

The best way to understand how this gauge independence appears is by 
noticing that when estimating $S$-matrix elements or expectation values 
of gauge independent magnitudes
one makes use of the 
quantum effective action on--shell, i.e., evaluated at those 
configurations that extremize it. According to Nielsen identities 
\cite{nielsen}, 
the variation of the quantum effective action due to changes in the 
functions that fix the gauge is
linear in the quantum--corrected equations of motion for the 
mean fields. It immediately follows that the on--shell quantum efective action 
 does not depend on the choice of the gauge--breaking term.
The mean fields do depend on the gauge--fixing, but this dependence 
exactly cancels out the explicit gauge--fixing dependence of the 
quantum effective action \cite{varios}.

In some situations one is interested in the
dynamical evolution of the fields rather than in the $S$-matrix
elements. We mention some examples: the analysis of 
phase transitions in field theory and condensed matter physics
\cite{zurek}, the
non-equilibrium aspects of the  
quark-gluon plasma \cite{plasma}, the quantum corrections to solitons
\cite{rocek}, and
the quantum corrections to the geometry in semiclassical
gravity, relevant in the early universe and black hole physics
\cite{review}. 
In all these cases the gauge--fixing dependence of the mean fields 
becomes a problem. This dependence does not contradict
the gauge--fixing independence of mean values of gauge
invariant operators, since these are in general non linear
functions of the fields, and it is not possible to compute
them directly from the mean values of the fields. 

One way to by--pass this difficulty is through a redefinition of the 
quantum effective action. Such is, for instance, the case of the 
Vilkovisky--De Witt formalism \cite{VdW}. Vilkovisky and De Witt 
 introduced a 
connection on the field--manifold and they used it in order to obtain a 
modified gauge--independent expression for the quantum effective action. 
However, despite of the gauge independence of the Vilkovisky--De Witt 
effective action, it depends on the choice of the connection on the 
configuration space, and there is not a universal criteria 
for the determination of such a connection, particularly in the case
of interacting fields \cite{odin}. In that sense, the 
Vilkovisky--De Witt definition is not a real solution for the problem
of the
gauge--fixing dependence of the mean fields.

Another possibility lies of course in considering effective potentials
or even actions
for gauge--invariant operators \cite{phi2}. In this framework, the 
issue of renormalization is not completely clear. It is also possible to construct an effective 
potential which is a  gauge--invariant function 
of a gauge--invariant order parameter.
This approach has been developed in equilibrium 
\cite{boyaetal} and also in non--equilibrium
situations \cite{heitmann}, for the abelian Higgs model. 
Extension to non--abelian
theories seems not straightforward. 

In this paper we present an alternative way to approach the subject, 
following the ideas exposed in \cite{Mazzitelli}, where 
the gauge--fixing problem was analyzed in the context of semiclassical gravity.
In that context, due to graviton corrections, the background
metric that solves the semiclassical Einstein equations is gauge--fixing dependent.
However, the metric can be ``measured'' by analyzing
the trajectory of a classical test particle, which plays the role of a 
classical device. The coupling of the particle to gravitons
compensates the gauge--fixing dependence of the metric, 
and the trajectory 
of the particle becomes gauge--fixing independent. 

In the present 
letter we will find similar results for a Yang-Mills theory.
Instead of introducing a new definition for the quantum effective action, 
we will consider a field theory involving a set of classical test fields 
interacting with a quantum background.
We study particularly the case of classical matter test fields 
interacting with a pure Yang--Mills environment. In that context, test fields 
could be interpreted as a classical device ``measuring'' physical quantities 
depending on the Yang--Mills degrees of freedom.

BRST--invariance of the 
classical effective action, when combined with the constraints imposed by 
the classical nature of the device, yields a set of ``generalized'' Nielsen 
identities expressing how does the dynamical evolution of the test fields 
depends on the choice of the gauge--fixing conditions. As it will be shown, 
the evolution of the classical device does not depend on the gauge--breaking 
term, even when the equations of motion of the quantum degrees of freedom 
do so. This problem has also been analyzed by Kazakov and Pronin \cite{Kasakov} using a 
different approach, and we will 
comment on this in the conclusions.

The paper is organized as follows. We first review how the Yang--Mills 
gauge system could be quantized according to the BRST--quantization 
procedure. Then, we derive the generalized Nielsen identity 
which governs, as we have mention, the gauge dependence of the quantum 
effective action. Finally, we explicitly show how the dynamics of 
the test fields is gauge--fixing independent.


\section{A generalized Nielsen identity}
\bigskip


We will consider a Yang-Mills theory coupled to classical matter fields.
Since we are interested in the dynamical evolution of the classical fields 
(considered as a classical device degrees of freedom), it will be useful 
to decompose the  action into two terms\footnote{Unless noted 
otherwise, we will use De Witt's condensed notation: indices will stand for 
all attributes of the fields and summation and integration over repeated 
indices will allways be understood. In addition, we will omit any kind of 
index if not strictly necessary. For example, we will write $A$ instead
of $A_{\mu}^a (x)$.}:
\[
\label{Action}
{\mathcal S}\left[\phi,A\right]={\mathcal S}_{\mbox{\scriptsize CD}}
\left[\phi\right]+{\mathcal S}_{\mbox{\scriptsize Q}}\left[\phi,A\right]
\]
The first term describes the free evolution of the classical device, while 
the second one describes the Yang--Mills environment and gives 
account of the 
interaction between both gauge and classical fields.
Fields $A$ can be freely thought as standing for gauge fields and for 
any other quantum matter fields, but since the consideration of other quantum 
degrees of freedom is straightforward, we will explicitly exclude them. 

Therefore, we assume that
\[
{\mathcal S}_{\mbox{\scriptsize Q}}={\mathcal S}_{\mbox{\scriptsize YM}}
\left[A\right]+\lambda{\mathcal S}_{\mbox{\scriptsize INT}}\left[\phi,A\right]
\label{decomp}
\]
${\mathcal S}_{\mbox{\scriptsize YM}}$ 
being the pure Yang--Mills free action and ${\mathcal S}_{\mbox{\scriptsize 
INT}}$ an interaction term. We have introduced the real number
$\lambda$ in order to parametrize the 
interaction strength, which we will assume to be weak.

Gauge--fixing is implemented by introducing into the classical action a 
gauge--breaking term depending on the gauge fields.
According to the BRST--quantization scheme, in order to construct this term 
it is useful to extend the field--configuration manifold by considering 
three new fields: the Fadeev--Popov fermionic ghost and anti--ghost fields, 
$\omega$ and  $\omega^{*}$, and the bosonic auxiliary Nakanishi--Lautrup fields, $h$, 
all of them carrying the group index \cite{BRST}.

Although the gauge--fixing term is not invariant under gauge 
transformations, it is required to be a BRST--invariant functional on the 
extended field--space. BRST--symmetry is defined in terms of a non--linear 
operator $s$, the so called Slavnov operator, according to
\[
\delta_{\mbox{\scriptsize BRST}}=\epsilon s
\]
where $\epsilon$ is an infinitesimal (global) fermionic parameter and the Slavnov operator 
acts on the fields in the following way:
\[
\label{eq:10}
s\phi_{\alpha} = i  \omega_{a} \left[t_{a} \right]_{\alpha\beta} \phi_{\beta}
\]
\[
\label{eq:12}
sA_{a\mu}= \left( D_{\mu} \omega\right)_{a} =\partial_{\mu} \omega_{a}+f_{abc}A_{b\mu}\omega_c
\]
\[
\label{eq:13}
s\omega_{a} = -\frac{1}{2}  f_{abc}\omega_{b} \omega_{c}
\]
\[
\label{eq:14}
s\omega^{*}_a = -h_{a}
\]
\[
\label{eq:15}
s h_{a} = 0
\]
Here the  $t_a$ are matrices carrying the fully reducible representation of the 
gauge group when acting on 
matter fields, $f_{abc}$ are the group structure constants and $D_{\mu}$ is, 
as usual, the covariant derivative in the adjoint representation.

If gauge--fixing at a classical level is imposed through conditions
\[
f\left[A\right]=0
\]
 the gauge--fixing action reads
\[
{\mathcal S}_{\mbox{\scriptsize GF}}\left[A,\omega,\omega^{*},h\right]=
h_af_a+\frac{1}{2}\alpha h_ah_a+\omega^{*}_a\frac{\delta f_a}{\delta A_{b\mu}}
 \left(D_{\mu}\omega\right)_b 
\]
where $\alpha$ is an arbitrary parameter {\footnote{Notice that  when integrated out in the path
integral, Nakanishi-Lautrup
 auxiliary fields $h$ reproduce the usual gauge-breaking term, quadratic in the 
functions fixing the gauge, $f$.}}.
The gauge--fixed action or classical effective action is, therefore,
\[
\label{effective}
{\mathcal I}\left[\phi, A,\omega,\omega^{*},h\right]={\mathcal S}_{\mbox{\scriptsize CD}}+{\mathcal S}_{\mbox{\scriptsize YM}}+\lambda{\mathcal S}_
{\mbox{\scriptsize INT}}+{\mathcal S}_{\mbox{\scriptsize GF}}
\]

BRST--transformations acting on the physical sector are simply 
reparametrizations of gauge transformations, implying that the 
gauge--invariant classical action ${\mathcal S}$ is also BRST--invariant. 
In fact, Fadeev--Popov ghost fields were introduced in order to realize 
the local gauge--invariance of the classical action as a global symmetry 
of the theory when defined on both the physical and unphysical
sectors.
 
Anti--ghost and Nakanishi--Lautrup fields, on the other hand, were 
introduced in order to ensure the Slavnov operator nilpotency, i.e.,
\[
s^2=0
\]
and 
to define the gauge--fixing condition as a Slavnov-variation. In fact,
it is straightforward to show that
\[
{\mathcal S}_{\mbox{\scriptsize GF}}=s\Psi\left[A,\omega,\omega^{*},h\right]
\]
where we have defined
\[
\Psi\left[A,\omega^{*},h\right]=\omega^{*}_a f_a+\frac{1}{2}\alpha
\omega^{*}_a h_a
\]
These properties imply the BRST--invariance of the
gauge--breaking term and, therefore, of the complete classical effective action.

The classical nature of the fields describing the device allows us, 
when calculating the connected vacuum persistence amplitude, to sum over 
all connected Feynman graphs excluding those diagrams involving loops 
with internal legs representing matter propagators. Its path integral 
representaton is, then, the usual one with integration just over gauge and 
unphysical fields: 
\[
{\mathcal W}\left[\phi,J\right] = -i \ln \int \left[d\chi\right] \ \exp{i\left\{ {\mathcal I}+J_n\chi_n\right\}}
\label{W}
\]
where, we are denoting by $\chi_n$ all fields 
but the classical ones, index $n$ running over all attributes 
of the fields, and $J$ are external sources coupled to the quantized fields.

The corresponding expression for the expectation value of an arbitrary
functional ${\mathcal F}$ defined on the extended field--space is
\[
\ll{\mathcal F}\rr = e^{-i{\mathcal W}} \int \left[d\chi\right] \ \exp{i\left\{ {\mathcal I}+J_n\chi_n\right\}} \ {\mathcal F}
\]

As can be seen from the equation above, the free action for the classical 
fields has no direct effect
on expectation values, since classical configurations are not
integrated and the purely classical contribution to ${\mathcal I}$, ${\mathcal S}_{\mbox{\scriptsize CD}}$, is
cancelled by normalization. Therefore, for simplicity, we will neglect
such contribution.

The quantum effective action is defined as the sum of all 1--particle 
irreducible graphs without loops with matter legs and, formally, 
it is given by the Legendre functional transform of ${\mathcal W}$,
\[
\Gamma\left[\phi,\overline{\chi}\right] = {\mathcal W} - J_n\overline{\chi}_n
\label{legendre}
\]

From Eq.\,(\ref{legendre}) it follows that $\overline{\chi}$ represents 
the expectation 
value of field $\chi$, since
\[
\overline{\chi}_n=\left(-\right)^{\chi}\frac{\delta {\mathcal W}}{\delta J_n}
\]
where functional derivatives are acting on the right and
$\left(-\right)^{\chi}=1$ if the field is bosonic and
$\left(-\right)^{\chi}=-1$ otherwise.

A path integral self--contained representation of the quantum effective 
action can be obtained by using
\[
\frac{\delta \Gamma}{\delta \overline{\chi}_n}=-J_n
\]
Introducing these expressions into Eq.\,(\ref{W}) it follows that
\[
\label{otra}
\Gamma = -i \ln \int \left[d \chi\right] \ \exp{i\left\{ {\mathcal I}-
\frac{\delta \Gamma}{\delta \overline{\chi}_n}
\left(\chi_n-\overline{\chi}_n\right)\right\}}
\]

Expectation values in terms of the quantum effective action are thus given by
\[
\ll{\mathcal F}\rr=e{-i\Gamma} \int \left[d\chi\right]  \ \exp{i\left\{ {\mathcal I}-\frac{\delta \Gamma}{\delta \overline{\chi}_n}\left(\chi_n-\overline{\chi}_n\right)\right\}}\ {\mathcal F} 
\label{F}
\]

We would like to compute the variation in the quantum effective action 
due to a change in the gauge--fixing conditions
$f \rightarrow f'=f+\Delta f$
and
$\alpha \rightarrow \alpha'=\alpha+\Delta \alpha$, 
or, in other words,
$\Psi \rightarrow \Psi'= \Psi+\Delta\Psi$.

Let us label the modified quantum effective action and expectation
values with a prime. Then, we have
\[
\label{otramas}
\Gamma' = -i \ln \int \left[ d\chi\right] \ \exp{i\left\{ {\mathcal I}+s
\Delta \Psi-\frac{\delta \Gamma}{\delta 
\overline{\chi}_n}\left(\chi_n-\overline{\chi}'_n\right)\right\}}
\]
where we have set the external sources to be the same as those
switched--on for the unchanged gauge condition, i.e.,
\[
\frac{\delta \Gamma'}{\delta \overline{\chi}'_n}=-J_n
\]
Substracting Eq.\,(\ref{otra}) from Eq.\,(\ref{otramas})
it follows that
\[
\label{dg}
\Delta \Gamma = -i \ln \ll \exp{i\left\{ s\Delta \Psi\right\}} \rr+\frac{\delta \Gamma}{\delta 
\overline{\chi}_n} \Delta \overline{\chi}_n
\]
where $ \Delta \overline{\chi} = \overline{\chi}' - \overline{\chi}$.

Up to first order in $\Delta\Psi$, Eq.\,(\ref{dg}) reduces to
\[
\Delta \Gamma = \ll s\Delta \Psi \rr+ \frac{\delta \Gamma}{\delta 
\overline{\chi}_n} \Delta \overline{\chi}_n
\]
where now $\Delta
  \overline{\chi}_n$ denotes the first order variation of the
  mean fields.

The so called Nielsen identity is simply the explicit form of 
this last equation. Performing a BRST--transformation in Eq.\,(\ref{F})
with ${\mathcal F}=\Psi$ 
and making use of the fact 
that BRST--transformations preserve volume in the configuration space (even 
when restricted to the quantum sector) it follows that
\[
\ll \Delta \Psi \rr = e^{-i\Gamma}\int \left[ d\chi\right] \ \exp{i\left\{ {\mathcal I}\left[\phi,\chi+\epsilon s \chi \right]-\frac{\delta \Gamma}{\delta \overline{\chi}_n}\left(\chi_n+\epsilon s \chi_n-\overline{\chi}_n\right)\right\}}
 \ \Delta \Psi\left[\chi+\epsilon s \chi\right]
\]

Now, using the BRST--invariance of the classical effective action,
\[
{\mathcal I}\left[\phi,\chi+\epsilon s \chi \right]={\mathcal I}
\left[\phi,\chi \right]-\epsilon \frac{\delta {\mathcal I}}{\delta 
\phi_{\alpha}} s\phi_{\alpha}
\]
we get
\[
\label{www}
\ll \Delta \Psi \rr = \ll \exp{i\left\{-\epsilon\frac{\delta \Gamma}{\delta \overline{\chi}_n}s\chi_n-\epsilon \frac{\delta {\mathcal I}}{\delta \phi_{\alpha}} s\phi_{\alpha}\right\}\left(\Delta \Psi+\epsilon s \Delta \Psi\right)}\rr
\]
Expanding the exponential in Eq.\,(\ref{www}), it follows that
\[
\Delta \Gamma=i\frac{\delta \Gamma}{\delta \overline{\chi}_n} \ll 
\left(s\chi_n\right) \Delta \Psi \rr+\frac{\delta \Gamma}{\delta 
\overline{\chi}_n} \Delta \overline{\chi}_n+i\ll \frac{\delta {\mathcal I}}{
\delta \phi_{\alpha}}\left(s \phi_{\alpha}\right) \Delta \Psi\rr
\label{nielsenid}
\]

This is the generalized Nielsen identity we have announced in the 
introduction. It controls the gauge dependence of the quantum effective 
action and it is the starting point for the demonstration of the 
gauge--fixing independence of any observable magnitude. In the absence of 
classical fields, the third term in the r.h.s of Eq.\,(\ref{nielsenid})
vanishes, and the usual Nielsen identity is recovered \cite{nielsen,varios}. 
In that particular
case (no classical fields) one can infer from the above identity
the gauge--fixing independence of spontaneous symmetry breaking,
Higgs masses
\cite{varios}, nucleation rates \cite{metaxas}, etc.

\section{Gauge--fixing independence of test fields}
\bigskip
\label{test}

We will now show that, up to first order in the interaction--strenght 
parameter $\lambda$, the  dynamics of the  classical fields turns out to be 
gauge--fixing independent when the mean values of the quantum fields 
are on--shell. That is, the explicit gauge--fixing dependence of
the mean values cancels out the explicit dependence of $\Delta\Gamma$.

Keeping only first order terms in $\lambda$ is justified by the fact
that we are assuming that the classical 
field
have no influence on
quantum mean values. 
Under these conditions, that we will refer as the {\it test field
conditions}, the quantum effective action has two main contributions: 
\[
\label{gggama}
\Gamma = \Gamma^{0}\left[\overline{\chi}\right] + \lambda
\Gamma^{1}\left[\phi,\overline{\chi}\right] + {\mathcal O}
\left(\lambda^2\right)
\]
where $\Gamma^{0}$ is the quantum effective action for the free 
Yang--Mills background and $\Gamma^{1}$ involves those terms 
describing the evolution of matter and its interaction with gauge
fields. In addition, the on--shell configurations of the quantized fields are
those obtained by solving the free Yang--Mills
equations of motion,
\[
\label{eqmotion}
\frac{\delta \Gamma^{0}}{\delta \overline{\chi}}\left[\overline{\chi}^{0}
\right]=0
\]
Eq.\,(\ref{gggama}) and Eq.\,(\ref{eqmotion}) imply that we can
freely add to the free on--shell mean fields
$\overline{\chi}^{0}$ an
arbitrary
 term linear in $\lambda$, since any first order correction
to these configurations has a second order effect 
in the final expression of the quantum effective action for matter, i.e.,
\[
\Gamma_{\text{{\scriptsize on--shell}}} = \Gamma\left[\phi,\overline{\chi}^{0}\right] \approx
\Gamma\left[\phi,\overline{\chi}^{0}+\lambda\xi \right]
\]
where we have denoted by  $\xi=\xi\left[\phi,\overline{\chi}^{0}\right]$ this
arbitrary additional term.
Accordingly,
\[
\Delta \Gamma_{\text{{\scriptsize on--shell}}} =\Delta \Gamma\left[\phi,\overline{\chi}^{0}\right] \approx
\Delta \Gamma\left[\phi,\overline{\chi}^{0}+\lambda\xi \right]
\]

The free Yang--Mills 
quantum effective action on--shell is gauge--independent. Moreover, the
variation of the mean fields due to changes in the gauge--fixing
conditions, $\Delta \overline{\chi}$, vanishes on--shell for the free case. 
Therefore, from Eq.\,(\ref{nielsenid}) we obtain
\[
\Delta \Gamma\left[\phi,\overline{\chi}^{0}+\lambda\xi \right] = 
i\lambda \left[\frac{\delta^2 \Gamma^{0}}{\delta
  \overline{\chi}_m \delta
  \overline{\chi}_n} \xi_m +
\frac{\delta \Gamma^{1}}{\delta
  \overline{\chi}_n}\right] \ll \left(s\chi_n\right) \Delta \Psi
\rr^{0}  +
i \lambda \ll \frac{\delta {\mathcal S}}{\delta \phi_{\alpha}}
\left(s
  \phi_{\alpha}\right) \Delta \Psi \rr^{0}
\]
Here derivatives of the
effective action are evaluated at $ {\chi}^{0}$ and
the superscript on the brackets means that expectation values are 
calculated neglecting the interaction between matter and gauge fields.

The functions $\xi$  still remain arbitrary. So,
it is possible to choose them such that they cancel
the r.h.s of the equation above.
With this choice it immediately follows that 
\[
\Delta \Gamma_{\text{{\scriptsize on--shell}}} =0
\]
showing that the dynamics of the test field is, up to first
order in $\lambda$, gauge--fixing independent.

\section{Conclusions}
\bigskip
\label{concl}

Being gauge--fixing dependent, the mean 
fields in Yang-Mills theories have no physical meaning. 
Therefore, to analyze quantum effects it is not enough to compute 
the effective action and solve the quantum corrected equations for the mean 
fields. It is also necessary to extract physical information from them.
In this 
paper we have shown that a way of doing this 
is to couple classical test fields to the Yang-Mills fields.
The dynamics of the test fields
is gauge--fixing independent, and therefore physically relevant.
These results are in tune with previous ones obtained in the context of 
semiclassical gravity \cite{Mazzitelli}.

In Ref. \cite{Kasakov}, the 
authors analyzed the same problem addressed here. Using a
Slavnov identity for the vacuum persistence amplitude
${\mathcal W}$, they showed that the dynamics of the 
classical fields is gauge--fixing independent in the 
low energy limit of power-counting renormalizable 
theories. In other words, they showed that within these reasonable
assumptions the  third term in the r.h.s of Eq.\,(\ref{nielsenid}) 
vanishes. 
In this paper we presented a different calculation 
based on Nielsen identity for the effective action.
In our approach we did not make any assumption
about  renormalizability.
This may be very useful for the analysis of the gauge--fixing
independence in semiclassical gravity, since 
the results are apparently dependent on the classical
device (see Refs. \cite{Mazzitelli} and \cite{Kasakov2}).    

We have not taken into account the divergences present in any field
theory. All our formal calculations should be understood  with a
regularization that does not break gauge symmetry. Moreover, when
considering classical fields, it may be necessary to add appropiate
counterterms to the action of the classical field. We will 
present some specific examples in a forthcoming publication.
\bigskip

{\bf Acknowledgements} 
This work was supported in part by CONICET, UBA,
and Agencia Nacional de Promoci\'on
Cient\'\i fica y Tecnol\'ogica. S.I. thanks FOSDIC foundation and
ARAGON foundation for support.

\end{document}